# FREELY FALLING OBSERVERS

Rainer Burghardt[*]

Keywords: Schwarzschild geometry, Lorentz transformation, accelerated systems, tidal forces.

Contents

We show that the field equations of the Schwarzschild geometry are invariant under passive Lorentz transformations to a freely falling system. We decompose the field equations with respect to the accelerated system and find that the force of gravity is not transformed away but dynamically compensated.

[*] e-mail: arg@onemail.at, home page: http://arg.at.tf/

# 1. INTRODUCTION

In Chapt. 2 we will discuss the behavior of the Schwarzschild geometry under passive Lorentz transformations. By an appropriate definition of covariant derivatives the field equations turn out to be invariant under these transformations. We make use of earlier results concerning methods of differential geometry [1].

In Chapt. 3 we will show that the field equations of the Schwarzschild geometry and also the subequations of the field equations are invariant under active Lorentz transformations. In the case of freely falling systems the force of gravity is not switched off by such a transformation but compensated by the acceleration of the falling observers. The remaining gravitational effects are the tidal forces.

In the next chapter we will reexamine the covariant field equations for these tidal forces and show that the geometrical meaning of these quantities are the second fundamental forms of a shrinking three-surface.

# 2. THE FREELY FALLING SYSTEM

Soon after the theory of general relativity was established by Einstein a wide discussion begun on the problem of the invariance properties of the theory. It was accepted by many searchers that the Christoffel symbols are not very useful to describe the physical content of the theory. A physically acceptable theory should be independent of the co-ordinates, but the Christoffel symbols don't have these properties. Using tetrads and the Ricci rotation coefficients has made some progress. Although the Ricci rotation coefficients are co-ordinate invariant they transform inhomogeneously under Lorentz transformations. Thus, any Lorentz transformation changes the physical content of these quantities. We will show how to avoid this problem.

In the Schwarzschild model the relative velocity of a freely falling observer with respect to a static observer has the radial component

$$v = -\sqrt{\frac{2M}{r}} \quad . \tag{2.1}$$

Operating with the generalized Lorentz transformation

$$L^1_{1'} = \alpha, \quad L^1_{4'} = -i\alpha v, \quad L^4_{1'} = i\alpha v, \quad L^4_{4'} = \alpha, \quad \alpha = 1/\sqrt{1-v^2} \tag{2.2}$$

on a covariant derivative we use a modified subsumption

$$\Phi_{m'||n'} = L^{m\ n}_{m'n'}\Phi_{m||n} = \left[\Phi_{m'|n'} - L^{s'}_{s}L^{s}_{m'|n'}\Phi_{s'}\right] - A_{n'm'}{}^{s'}\Phi_{s'}, \quad A_{n'm'}{}^{s'} = L^{n\ m\ s'}_{n'm's}A_{nm}{}^{s} \qquad (2.3)$$

and we define a new covariant derivative

$$\Phi_{m'\underset{1}{||}n'} = \Phi_{m'|n'} - L_{n'm'}{}^{s'}\Phi_{s'}, \quad L_{n'm'}{}^{s'} = L^{s'}_{s}L^{s}_{m'|n'} \qquad (2.4)$$

which fits neatly the graded derivatives proposed in [1]. It reduces to the ordinary partial derivative only for special reference systems in a similar way, as the ordinary covariant derivative in flat space reduces to the partial one, if a Cartesian system is chosen. This derivative leaves invariant the autoparallelism of the unit vectors

$$m_n = \{1,0,0,0\}, \quad b_n = \{0,1,0,0\}, \quad c_n = \{0,0,1,0\}, \quad u_n = \{0,0,0,1\}$$

as

$$m_{m'\underset{1}{||}n'} = m_{m'|n'} - L_{n'm'}{}^{s'}m_{s'} = 0, \quad b_{m'\underset{2}{||}n'} = b_{m'|n'} - L_{n'm'}{}^{s'}b_{s'} = 0, \quad c_{m'\underset{3}{||}n'} = c_{m'|n'} - \left(L_{n'm'}{}^{s'} + B_{n'm'}{}^{s'}\right)c_{s'} = 0$$

$$u_{m'\underset{4}{||}n'} = u_{m'|n'} - \left(L_{n'm'}{}^{s'} + B_{n'm'}{}^{s'} + C_{n'm'}{}^{s'}\right)u_{s'} = u_{m'|n'} - L_{n'm'}{}^{s'}u_{s'} = 0$$

By using (2.2) we obtain the above connexion coefficients[1] from

$$A_{mn}{}^{s} = B_{mn}{}^{s} + C_{mn}{}^{s} + E_{mn}{}^{s}, \quad B_{mn}{}^{s} = b_m B_n b^s - b_m b_n B^s$$
$$C_{mn}{}^{s} = c_m C_n c^s - c_m c_n C^s, \quad E_{mn}{}^{s} = -\left[u_m E_n u^s - u_m u_n E^s\right] \qquad (2.5)$$

where the quantities

$$B_n = \left\{\frac{a}{r},0,0,0\right\}, \quad C_n = \left\{\frac{a}{r},\frac{1}{r}\cot\vartheta,0,0\right\}, \quad E_n = \left\{\frac{1}{\rho}\frac{v}{a},0,0,0\right\} \qquad (2.6)$$

are the curvatures of the physical surface, $a = \sqrt{1 - 2M/r}$ and $\rho(r) = \sqrt{2r^3/M}$ the curvature vector of the Schwarzschild parabola. Operating with (2.2) on (2.6) we obtain the components of these quantities in the freely falling system

---

[1] These quantities and the physical surface has been discussed in detail in paper [1]

$$B_{n'} = \left\{\frac{1}{r}, 0, 0, -iv\frac{1}{r}\right\}, \quad C_{n'} = \left\{\frac{1}{r}, \frac{1}{r}\cot\vartheta, 0, -iv\frac{1}{r}\right\}, \quad E_{n'} = \left\{\alpha\frac{1}{\rho}\frac{v}{a}, 0, 0, -i\alpha v\frac{1}{\rho}\frac{v}{a}\right\}. \quad (2.7)$$

Since the Ricci tensor is invariant under the Lorentz transformation $R_{m'n'} = L_{m'n'}^{m\,n} R_{mn} = 0$ we get with

$$R_{m'n'} = A_{m'n'\,\|\,s'}^{s'} - A_{n'\|m'}^{s'} - A_{r'm'}^{s'} A_{s'n'}^{r'} + A_{m'n'}^{s'} A_{s'} \quad (2.8)$$

the same equations that we derived in [1] for the static system with primed indices

$$R_{n'm'} = -\left[B_{m'\|n'} + B_{m'}B_{n'}\right] - b_{m'}b_{n'}\left[B^{r'}_{\|r'} + B^{r'}B_{r'}\right] - \left[C_{m'\|n'} + C_{m'}C_{n'}\right] - c_{m'}c_{n'}\left[C^{r'}_{\|r'} + C^{r'}C_{r'}\right] +$$
$$+ \left[E_{m'\|n'} + E_{m'}E_{n'}\right] + u_{m'}u_{n'}\left[E^{r'}_{\|r'} + E^{r'}E_{r'}\right] = 0, \quad B_{[m'\|n']} = 0, \quad C_{[m'\|n']} = 0, \quad E_{[m'\|n']} = 0 \quad (2.9)$$

The connexion coefficients $A_{nm}^{\ s}$ refer to an invariant geometrical structure. They are expressed by the curvatures of the slices of the physical surface, namely $|B| = 1/r$, $|C| = 1/r\sin\vartheta$, $|E| = 1/\rho a$ and behave like a tensor under Lorentz transformations. The geometrical content of the theory is not altered by a rotation of the reference system in the tangent space. The primed equations stand for the prediction that a freely falling observer makes for the physics of the static observers.

## 3. THE ACTIVE TRANSFORMATION

In the last chapter we have demonstrated that the field equations of the Schwarzschild geometry and also its subequations are invariant under passive Lorentz transformations. Now we will analyze the consequences by interpreting (2.2) as an active transformation. There are two reference systems in use, the static one

$$m_{s'} = \{\alpha_s, 0, 0, -i\alpha_s v_s\}, \quad b_{s'} \doteq b_s, \quad c_{s'} \doteq c_s, \quad u_{s'} = \{i\alpha_s v_s, 0, 0, \alpha_s\} \quad (3.1)$$

and the accelerated one

$$'m_{s'} = \{1,0,0,0\}, \quad 'b_{s'} = b_{s'}, \quad 'c_{s'} = c_{s'}, \quad 'u_{s'} = \{0,0,0,1\}. \tag{3.2}$$

In Chapt. 2 we have decomposed the field equations with respect to the reference system (3.1). To get the physics measured by the observers in their own system we decompose the field equations with respect to (3.2). To begin with, we calculate the quantity $L_{n'm'}{}^{s'}$ resulting from the fact that a transformation to an accelerated system has non-constant parameters:

$$L_{n'm'}{}^{s'} = m_{n'} Q_{m'} m^{s'} - m_{n'} m_{m'} Q^{s'}, \quad Q_{m'} = \left\{0,0,0,-\frac{i}{\rho}\right\} \tag{3.3}$$

or

$$L_{n'm'}{}^{s'} = Q_{n'm'}{}^{s'} + G_{n'm'}{}^{s'}, \quad G_{m'} = \left\{\alpha \frac{1}{\rho}\frac{v}{a}, 0, 0, -i\alpha v \frac{1}{\rho}\frac{v}{a}\right\}$$

$$Q_{n'm'}{}^{s'} = 'm_{n'} Q_{m'} 'm^{s'} - 'm_{n'} 'm_{m'} Q^{s'}, \quad G_{n'm'}{}^{s'} = u_{n'} G_{m'} u^{s'} - u_{n'} u_{m'} G^{s'} \tag{3.4}$$

The new quantity Q is the radial tidal force acting on the freely falling observers and the quantity $G_m = L_m^{m'} G_{m'}$ the acceleration of the freely falling system measured by the observers of the static system. As

$$'u_{m'\|s'} \, 'u^{s'} = -G_{m'}$$

the acceleration G and the gravitational force E occur with the opposite sign in the theory. Since they also have the same value they cancel out in the theory. Using the total derivative

$$'u_{m'\|s'} \, 'u^{s'} = -\left(L_{s'm'}{}^{r'} + G_{s'm'}{}^{r'}\right) 'u_{r'} \, 'u^{s'} = -(G_{m'} - E_{m'}) = 0 \tag{3.5}$$

we find that the freely falling observers can experience no acceleration. The last equation shows that gravitation cannot be transformed away by a suitable Lorentz transformation but is dynamically *compensated*. This is outlined by the fact that both quantities have different roots. The force of gravity has its origin in the curvature of the Schwarzschild parabola while the acceleration is derived from a 4-bein structure in the tangent space:

$$G_{m'} = \frac{1}{\alpha}\alpha_{|m'} = \frac{1}{ch\chi}(ch\chi)_{|m'}, \quad E_{m'} = -\frac{1}{a}a_{|m'} = -\frac{1}{\cos\varepsilon}(\cos\varepsilon)_{|m'}.$$

Herein are $\chi$ the angles of rotation in the tangent space and $\varepsilon$ the angles of ascent of the Schwarzschild parabola. The fact that $1/\alpha$ and a are numerically equal provides the compensation of the force of gravity by the acceleration term. How to switch on and off gravitation by non-Lorentzian transformations has been discussed in [2, 3, 4].

Transforming in the Ricci tensor the connexion coefficients inhomogeneously, the condition

$$L_{m'n'\underset{1}{\|s'}}{}^{s'} - L_{s'n'\underset{1}{\|m'}}{}^{s'} - L_{s'm'}{}^{r'}L_{r'n'}{}^{s'} + L_{m'n'}{}^{s'}L_{r's'}{}^{r'} + 2A_{[m's']}{}^{r'}L_{r'n'}{}^{s'} = 0 \qquad (3.6)$$

must be satisfied to keep the Ricci tensor Lorentz invariant. Equs. (3.6) are the field equations for the new quantities incorporated into the theory by the Lorentz transformation. They describe an additional structure in the tangent space of the physical surface. They decouple from the field equations, while the remaining quantities describe the geometrical structure of the physical surface. For the Schwarzschild geometry this relation has the form

$$\left[Q_{n'|m'} - Q_{n'|s'} {}'m^{s'} {}'m_{m'} + Q_{n'}Q_{m'}\right] + {}'m_{n'} {}'m_{m'}\left[Q^{s'}{}_{|s'} + Q^{s}Q_{s'}\right] +$$
$$+ \left[G_{n'\underset{1}{\|m'}} - G_{n'}G_{m'}\right] + u_{n'}u_{m'}\left[G^{s'}{}_{\underset{1}{\|s'}} - G^{s}G_{s'}\right] = 0 \qquad (3.7)$$

If we take into account the relation (3.7) and G=E we are able to compensate the force of gravity with the radial acceleration in the field equations. The best representation of these new equations is the five-dimensional one, developed in paper [1]

$$B_{n'\underset{2}{\|\|m'}} + B_{n'}B_{m'} = 0, \quad C_{n'\underset{3}{\|\|m'}} + C_{n'}C_{m'} = 0$$
$$B^{s'}{}_{\underset{2}{\|\|s'}} + B^{s'}B_{s'} = 0, \quad C^{s'}{}_{\underset{3}{\|\|s'}} + C^{s'}C_{s'} = 0, \quad Q^{s'}{}_{\|\|s'} = 0 \qquad (3.8)$$

Herein the five-dimensional covariant derivative is used with the additional components $M_{a'b'}{}^{c'} = m_{a'}M_{b'}m^{c'} - m_{a'}m_{b'}M^{c'}$, $M_{b'} = \left\{\dfrac{1}{\rho}, 0, 0, 0, 0\right\}$, $a' = 0', 1', ..., 4'$. The extra dimension is the 0-direction and M is the curvature of the physical surface in its rigging direction. As the first two equation of (3.8) contain tidal forces too we will develop equations describing the tidal forces uniquely in the next chapter.

# 4. TIDAL FORCES

The relations we have derived in the last chapter improve our insight into the invariance behavior of gravitation theory under active Lorentz transformation and the mechanism how fields are created and compensated dynamically. But these relations do not present the physical substance of the theory in a satisfactory way. The tidal forces

$$B_{4'} = \frac{-iv}{r}, \quad C_{4'} = \frac{-iv}{r}, \quad Q_{4'} = \frac{-i}{\rho}, \qquad (4.1)$$

stretching and squeezing a freely falling observer on his way to the center of attraction, are scattered to several subequations. We will rearrange the field equations for the sake of getting only one set of subequations for the tidal forces. From now on we drop the primes on the indices. Defining the non vanishing-components of a symmetric quantity Q by

$$Q_{11} = -\frac{i}{\rho}, \quad Q_{22} = -\frac{iv}{r}, \quad Q_{33} = -\frac{iv}{r} \qquad (4.2)$$

we express the tidal forces by a second rank tensor [5,6]. As

$$Q_{mn} = {'u}_{m\|n}, \quad Q_{mn}{'u}^m = 0, \quad Q_{mn}{'u}^n = 0, \qquad (4.3)$$

the components of the tensor Q are the second fundamental forms of three-dimensional space-like surfaces shrinking on the observers' ways to the central mass. We perform an inhomogeneous transformation of the connexion coefficients by use of (2.2) and split the new connexion coefficients into a space-like part *A and a time-like part containing the second fundamental forms

$$A_{mn}{}^s = {}^*A_{mn}{}^s + Q_m{}^s{'u}_n - Q_{mn}{'u}^s. \qquad (4.4)$$

The three dimensional space-like covariant derivative reads as

$$\begin{array}{c}\Phi_{\alpha\wedge\beta} = \Phi_{\alpha|\beta} - A_{\beta\alpha}{}^\gamma \Phi_\gamma, \quad A_{\beta\alpha}{}^\gamma = B_{\beta\alpha}{}^\gamma + C_{\beta\alpha}{}^\gamma \\ B_{\beta\alpha}{}^\gamma = b_\beta B_\alpha b^\gamma - b_\beta b_\alpha B^\gamma, \quad C_{\beta\alpha}{}^\gamma = c_\beta C_\alpha c^\gamma - c_\beta c_\alpha C^\gamma\end{array}, \qquad (4.5)$$

where Greek indices are running from 1 to 3. Since

$$B_\alpha = \left\{\frac{1}{r},0,0\right\}, \quad C_\alpha = \left\{\frac{1}{r},\frac{1}{r}\cot\vartheta,0\right\} \tag{4.6}$$

the freely falling system pretends a flat three-dimensional geometry. But one has to bear in mind that the curvatures are invariant quantities, and that the first components of the curvatures fully written out are

$$B_1 = \alpha a \frac{1}{r}, \quad C_1 = \alpha a \frac{1}{r}. \tag{4.7}$$

The two factors α and a have different roots but compensate. The field equations for these quantities read as

$$B_{\alpha|\beta} + B_\alpha B_\beta = 0, \quad B^\gamma{}_{|\gamma} + B^\gamma B_\gamma = 0, \quad C_{\alpha|\beta} + B_{\beta\alpha}{}^\gamma C_\gamma + C_\alpha C_\gamma = 0, \quad C^\gamma{}_{|\gamma} + B_\gamma C^\gamma + C_\gamma C^\gamma = 0,$$
$$B_{\beta|4} + Q_{22} B_\beta = 0, \quad C_{\beta|4} + Q_{33} C_\beta = 0 \tag{4.8}$$

where the last two equations describe the change of the spatial curvature measured with the proper time by the freely falling observers in falling towards the central mass. The field equations also contain relations for the tidal forces

$$\begin{aligned}
R_{\underline{mn}} &= -\left[Q_{mn\wedge s}'u^s + Q_{mn}Q_s^s\right] = 0 \\
R_{\underline{mn}}'u^n &= -\left[Q_s{}^s{}_{\wedge m} - Q_m{}^s{}_{\wedge s}\right] = 0 \\
R_{mn}'u^m{}'u^n &= -\left[Q_s{}^s{}_{\wedge m}'u^m - Q_{rs}Q^{rs}\right] = 0
\end{aligned} \tag{4.9}$$

The underlined indices denote the space-like projections with respect to the falling observers. The second equation of (4.9) is the contracted equation of Codazzi for the second fundamental forms.

## 5. CONCLUSIONS

In this paper we have shown that the Schwarzschild model is invariant under Lorentz transformations. This result can be generalized for other gravitational models. To interpret the Lorentz transformation as an active one we decompose the field equations with respect to the Lorentz rotated 4-bein. Since for accelerated systems the Lorentz transformation has non-constant parameters, new quantities are dynamically created due to the motion of these systems. For the case of the freely falling systems in the Schwarzschild geometry the force of gravitation is compensated by the acceleration of the freely falling observers. Thus, these observers can measure no gravitational force, but tidal forces act on them. As it is more convenient to represent these quantities as second rank tensors than as vectors the picture of the theory is changed drastically and the invariance under active Lorentz transformations is hidden.